\documentclass{article} 
\usepackage{iclr_original,times}

\usepackage{amsmath,amsfonts,bm}

\def\eqref#1{equation~\ref{#1}}

\def\1{\bm{1}}

\DeclareMathAlphabet{\mathsfit}{\encodingdefault}{\sfdefault}{m}{sl}
\SetMathAlphabet{\mathsfit}{bold}{\encodingdefault}{\sfdefault}{bx}{n}

\usepackage{graphicx}
\usepackage{hyperref}
\usepackage{url}
\usepackage{subcaption}
\usepackage{wrapfig}
\usepackage{enumitem}

\newcommand\blfootnote[1]{%
  \begingroup
  \renewcommand\thefootnote{}\footnote{#1}%
  \addtocounter{footnote}{-1}%
  \endgroup
}
\title{Rethinking Machine Learning Model Evaluation in Pathology}

\author{\parbox{\linewidth}{\centering Syed Ashar Javed\textsuperscript{*}, Dinkar Juyal\textsuperscript{*}, Shreya Chakraborty\textsuperscript{*}, Zahil Shanis\textsuperscript{*}, Harsha Pokkalla, Aaditya Prakash\\~\\
{\normalfont PathAI Inc., Boston}\\
{\normalfont \texttt{\{firstname.lastname\}@pathai.com}} \\
}}

\iclrfinalcopy 
\begin{document}

\maketitle
\begin{abstract}

Machine Learning has been applied to pathology images in research and clinical practice with promising outcomes. However, standard ML models often lack the rigorous evaluation required for clinical decisions. Machine learning techniques for natural images are ill-equipped to deal with pathology images that are significantly large and noisy, require expensive labeling, are hard to interpret, and are susceptible to spurious correlations. We propose a set of practical guidelines for ML evaluation in pathology that address the above concerns. The paper includes measures for setting up the evaluation framework, effectively dealing with variability in labels, and a recommended suite of tests to address issues related to domain shift, robustness, and confounding variables. We hope that the proposed framework will bridge the gap between ML researchers and domain experts, leading to wider adoption of ML techniques in pathology and improving patient outcomes.
\end{abstract}

\section{Introduction} \label{introduction}

Pathology is often treated as a ground-truth for many serious diseases and a pathologist's diagnosis is critical for a wide variety of tasks, including  drug development and clinical diagnostics. The typical workflows of a pathologist are complicated, subjective and limited by what humans can evaluate. The use of machine learning in pathology has begun to transform the field with  applications like clinical decision support tools \citep{campanella2019clinical}, improving prognostic utility over manual reads \citep{taylor2021machine}, automating slide scoring in clinical trials \citep{glass2021machine}, novel biomarker discovery \citep{echle2021deep}, and understanding tumor pathology \citep{jiang2020emerging, diao2021human}\blfootnote{\textsuperscript{*}equal contributions}.

These models require thorough validation and verification since they directly impact patient lives and their evaluation requires a careful look at the setting in which they will be deployed. Pathology images also differ from natural images in various ways - they are massive in resolution (billions of pixels) with each magnification-level containing markedly different information \citep{komura2018machine}, have different equivariances from natural images \citep{veeling2018rotation} and contain stain variations which can confound models \citep{yagi2011color, madabhushi2016image, tellez2019quantifying}. In addition to that, exhaustive labeling of these images is prohibitive as they require expert annotations and are time-consuming. Finally, building ML models for these problems requires understanding of biological causal structures to deal with spurious features \citep{castro2020causality}. These differences necessitate model evaluation guidelines specific to pathology problems to prevent life threatening model failures.

In machine learning, dataset creation, evaluation setup, choice of metrics, and the data splits used for training, validation and testing have been areas of study in themselves, and the research community depends on agreed standards for evaluating novel methods and their utility in the real world. Previous work in medical imaging has highlighted problems in transferring these evaluation methodologies, making models brittle for deployment in real-world setting \citep{park2018methodologic, england2019artificial, varoquaux2021failed, reinke2021common}. In this work, we propose a guideline on ML model evaluation for pathology which go beyond standard ML metrics. We define how to setup the evaluation framework, create the test-set, acquire labels for evaluation and finally suggest a set of evaluation metrics which are aimed towards real-world deployment.
\vspace{-5pt}
\section{Guidelines for Model Evaluation}
\begin{figure}
  \begin{center}
    \includegraphics[scale=0.23]{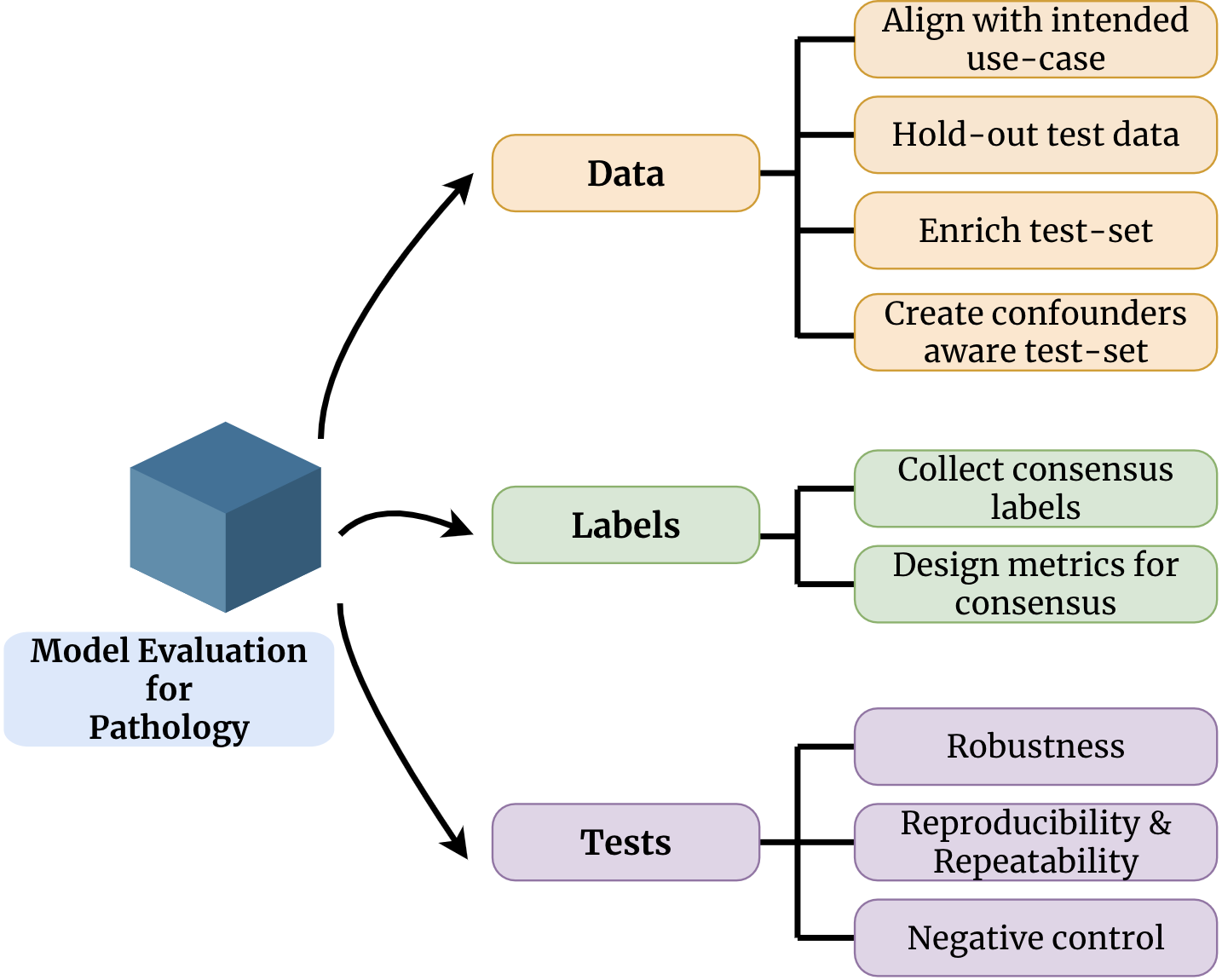}
  \end{center}
  \caption{Guidelines for Model Evaluation}
  \label{fig:pipeline}
\end{figure}

We divide model evaluation into three stages: evaluation setup, label collection, and evaluation metrics as shown in Figure \ref{fig:pipeline}. We provide pathology-specific recommendations for each of them in the following sections.

\subsection{Evaluation Setup}
\subsubsection{Alignment of evaluation experiments with model’s intended use}

The applications of ML in pathology vary by their intended use, such as drug discovery, diagnostics, prognostics, decision-support, or triage. Therefore, their evaluation setup and metrics must account for the targeted use-case.
For example, a clinical decision support tool is used by pathologists to look at predictions generated by the model to refine their diagnosis. The evaluation setup must measure the change in pathologist performance when assisted by the model instead of comparing model predictions against that of the pathologist.

Another example is the use of ML in building a triage tool for screening patients. In oncology, ML can be used to filter out obviously benign cases so that the pathologists prioritize malignant cases, thus improving patient care. In such scenarios, instead of using common metrics like accuracy or AUROC, one should use Precision@K\%Recall due to the low tolerance for a false negative.

\subsubsection{Enriched and confounder-aware test set creation}
Pathology samples come with biological and other metadata like patient-related (age, gender, ethnicity), image-related  (scanner type, image format, magnification), specimen \& stain related (tissue collection type, thickness of specimen, staining protocol), and clinical (disease stage, prescribed therapy, omics data). Low prevalence of certain metadata values coupled with small, imbalanced datasets in pathology often lead to overoptimistic results if the test set composition does not account for smaller substrata.
This hidden stratification problem \citep{oakden2020hidden} can result in the model not working well for patients belonging to a hidden substrata, which in turn deteriorates trust in such tools. Another stratum which needs more representation is samples close to the decision boundary where significant ambiguity in labels exist. This occurs since labels in pathology represent a biological process which is continuous. For example, along with benign and malignant there exist borderline cases which might be under-represented if real-world data distribution is replicated in test dataset.

A related problem is the presence of confounders in the test set which are spuriously predictive of the label. For example, in pancreatic cancer, \citep{kather2020pan} show that KRAS mutation can be predicted from a pathology image. However, KRAS mutation is almost exclusively present in only one sub-type of pancreatic cancer (pancreatic ductal adenocarcinoma), thus becoming a strong biological confounder \citep{waters2018kras}. Therefore, even a well performing mutation prediction model might be relying on pancreatic sub-type features instead of the KRAS mutation features. The fact that these confounders can also be partial, non-biological and specific to the dataset, further complicates matters \citep{badgeley2019deep, larrazabal2020gender, zech2018variable}.

To ensure that the test-set is reliable and the model is predicting the variable of interest and not a confounder, we offer the following suggestions: 

\begin{itemize}[leftmargin=.1in] 
  \item Curate test sets such that they have enough representation of any sub-strata of data that we care about or that could be a confounder \citep{seyyed2020chexclusion}. This can be done through enrichment (oversampling) of rare substrata as mentioned in this FDA guide \citep{foodguidance}. However practitioners should be aware of selection bias due to test-set enrichment \citep{yu2020one}.
  \item Incorporate stratified evaluation metrics as opposed to a single test-set wide global metric. 
  For example, ensuring consistent performance across patient demographics or disease severity.
  \item Conduct a correlation check on all metadata values present in the dataset to see if any of them correlate highly with label values for the task of interest. For example, if all images positive for prostate cancer come from the same hospital, the model can learn to predict the hospital instead of the cancer, making it a potential confounder.
\end{itemize}

\subsubsection{External hold-out test set}
Curated datasets are useful for research, but often suffer from the problem of a homogeneous test-set. The approach of collecting data from a single source and splitting it into train-val-test sets results in inflated performance since the models might not generalize to distributional shifts (like images from a new hospital) \citep{tellez2019quantifying, campanella2019clinical}. Many popular pathology datasets \citep{borkowski2019lung, hosseini2019atlas, karimi2019deep} and studies \citep{shao2021transmil, sudharshan2019multiple} do not have externally held-out test sets from a new site which prevent an evaluation of the generalizability. We suggest using multiple held-out patient cohorts from different medical sites and demographics while constructing the test set \citep{koh2021wilds}. We also suggest checking for test data leakage in the form of patient samples being present across multiple splits. Finally, we urge the community to make such a testing setup the norm for model evaluation in pathology.
\vspace{-5pt}

Figure illustrates the discordance among pathologists while grading NASH, an aggressive form of non-alcoholic fatty liver disease increasingly being seen as an epidemic in the US. In this case, NASH severity is found by performing a liver biopsy, viewing the extracted tissue through a microscope and observing signs of disease progression such as presence of fat, different kinds of inflammation etc. 

What is most telling in this figure is the extremely low agreement among doctors while predicting disease progression even after using the most reliable way to do so, which is viewing the tissue extracted from the biopsy. The numbers on the left column denote the Kappa score for different morphological indicators of NASH (N denotes the number of patients), when the first study was done in 2005. The right column shows the agreement scores for the same from a 2020 study; there has been little or no change in these agreement scores among human experts even after 15 years.


\subsection{Evaluation Labels}

\begin{figure}
    \subfloat[Figure illustrates the discordance among human experts while grading NASH from pathology images. NASH is an aggressive form of non-alcoholic fatty liver disease increasingly being seen as an epidemic in the US. Kappa score (agreement score) among pathologists for different morphological indicators of NASH are shown 15 years apart; there has been no improvement in pathologist consensus.]
    {
    \label{fig:score_discordance} \includegraphics[width=.5\textwidth, height=.25\textwidth]{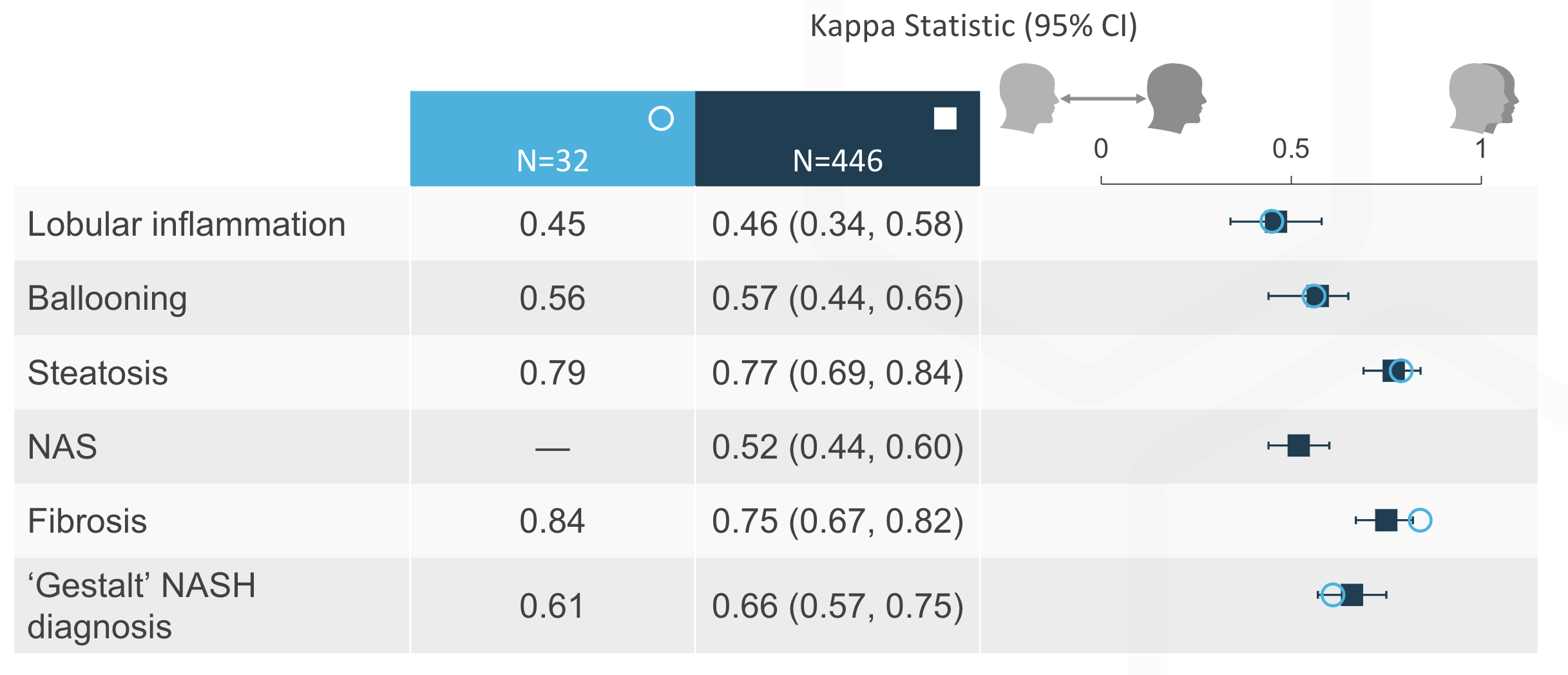}}
    \hspace{0.7cm}
    \subfloat[Variations in  pathology images across scanners and lab sites. The first row shows the same slide scanned with different scanners with visually perceptible differences. The second row shows slides collected from the same scanner but different lab sites, which differ visually due to staining protocols, choice of reagents, method of specimen preparation and so on. Models struggle to generalize to these variations]{\label{fig:stain_variation}
    \includegraphics[width=.4\textwidth, height=.25\textwidth]{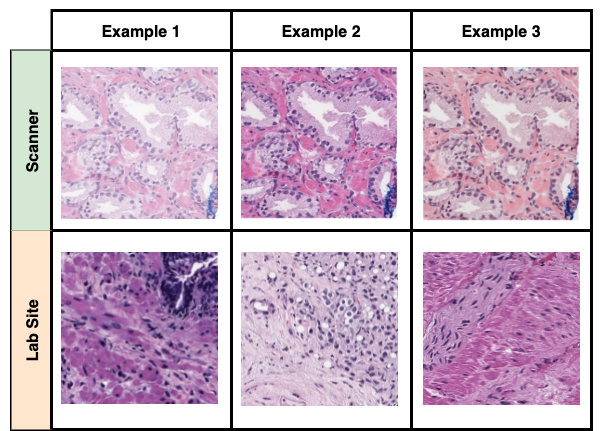}}
    \centering
    \caption{}
    \label{fig:my_label}
\end{figure}

It is possible to get fast and cheap labels using crowdsourced tools for natural images, and the correctness of these labels is easily verifiable. In contrast, it is a lot harder to collect ground-truth labels for pathology. Pathology images have billions of pixels, millions of cells, and thousands of regions of interest, making label collection extremely resource-intensive. Annotators (typically certified pathologists) are in short supply and per unit cost of annotations is order of magnitude higher than natural images. Annotators also have their own biases \citep{10.5858/arpa.2016-0386-RA}.

A more insidious problem is of the high `inter-annotator variability' \citep{carrasco2021ai} as shown in Figure \ref{fig:score_discordance},  and `intra-annotator variability' \citep{intra-rater-variability, DAVISON20201322}: annotators disagree not only with each other but also with their past selves. These variabilities exist at different stages of label generation - they can originate from inherent ambiguity in the labeling system, different interpretations of the same labeling scheme by different experts, and the differences in the expertise of these annotators \citep{ozkan2016interobserver}. The threshold for acceptable variability is subjective, and depends on factors such as the particular organ and disease. 

We propose the following guidelines while collecting labels:
\begin{itemize}[leftmargin=.1in] 

    \item Collect multiple annotations per image and measure the inter-annotator agreement. If possible, collect multiple annotations on the same image from the same annotator at adequately-spaced timepoints (called `washout period') to measure intra-annotator variability. If these variabilities are high, the ML task might not be viable.
    \item Aggregate the labels to a single consensus if the intra and inter-annotator variabilities are low.
    Instead of comparing model performance with a single pathologist \citep{liu2017detecting} or consensus, measure the agreement of the model with the consensus and judge it against inter-annotator consensus. This is more realistic and appreciates the difficulty of the task as judged by human's inability to do it consistently \citep{raciti2020novel}. Kappa (Linear, Quadratic) and Intraclass Correlation Coefficient(ICC) are popularly used to measure single and multi-rater agreement \cite{bulten2022artificial, koo2016guideline}.
\end{itemize}

\subsection{Evaluation Metrics}

\subsubsection{Reproducibility, Repeatability, \& Robustness tests}
%
Pathology images are heterogeneous with regard to staining, tissue thickness, and are prone to variations during tissue processing, cutting, staining, and digitization (Figure \ref{fig:stain_variation}).
These variations cause minor visually perceptible changes to the image, but can drastically change the model prediction.
Lack of robustness to these perturbations poses a major challenge to the wider adoption and use of ML in pathology, especially in a clinical setting \citep{stresstesting}. 
Accuracy based assessment of models, although necessary, is not sufficient for capturing model generalization. Following tests must be incorporated into the model evaluation framework to assess model's robustness to these variations:
\begin{itemize}[leftmargin=.1in]
    \item Reproducibility Tests:
    Models must be tested under known sources of variations and evaluated for consistency and accuracy in performance. This includes testing by explicitly changing external variations such as staining \& scanning, variations associated with hardware (non-determinism in GPU) and software packages. Models in pathology are usually trained on smaller patches of the image due to their large size \citep{shao2021transmil, taylor2021machine}. Stochasticity introduced due to sampling the patches can lead to inconsistent predictions and robustness to this variability must also be reported.
    \item Repeatability Tests: Repeatability expresses the precision under the same operating conditions over a short interval of time \citep{fda_robustness}. This test captures unknown \& unaccountable variations that are not covered by reproducibility tests. Consistency in model predictions must be measured and reported by rescanning \& rerunning inference on the same image multiple times, keeping all known sources of variations constant. 
    \item Robustness to Perturbations: Creating test sets that capture all possible real-world variations of images is not practical. An alternative can be to evaluate robustness on synthetically generated perturbations inspired by real world variations to image samples, by measuring consistency of predictions across these perturbations \citep{hendrycks2019benchmarking, stresstesting}. Models that are robust to synthetic variations have been shown to be robust to natural variations \citep{faryna2021tailoring}.
\end{itemize}

\subsubsection{Negative Control Tests}

Cross domain differences in pathology images collected from different sources may confound the discovery of explanatory variables and lead to poor generalization. This phenomenon, called `batch effects' \citep{goh2017batch}, can lead to the model learning spurious features.
Existing literature on batch effects show that models can predict patient age, sex, race, time of slide preparation, site, and scanner \citep{schmitt2021_batch, howard2020_batch}. 


Although creating confounder-aware test sets can mitigate some of these issues, partial and hidden confounders may go unnoticed and affect model performance. Another approach to tackle this issue is to perform negative control tests which involve running experiments under conditions in which the model is not expected to produce correct predictions \citep{negativecontrol}. We suggest the following strategies for designing negative control experiments for ML in pathology:
\begin{itemize}[leftmargin=.1in]
    \item Train with regions of the image which have no discriminative information. For example, a model trained to predict cancer using only background (non-tissue area) should perform poorly.
    \item Train without regions which are essential for predicting the label. For example, models for predicting cancer should perform poorly when trained and deployed on patches sampled from non-cancer regions.
\end{itemize}
Failure of a negative control test signals the presence of confounders and biases that create spurious correlations. These tests are not designed to detect all types of confounders, but they help detect the prominent ones and take appropriate actions.


\section{Conclusion}

Machine learning for pathology requires different evaluation standards due to differences in image characteristics and clinical use-case. Existing datasets and evaluation metrics prevalent within the ML community need to be rethought to fit pathology applications. We have observed that there exist two distinct communities, one of ML researchers focused on the development of novel methods and the other of computational pathology researchers focused on exploring and utilizing the power of ML in healthcare, and model evaluation differs between them materially. We argue that domain-specific knowledge from pathology concerning label reliability, test set enrichment, and biological confounders should permeate into ML research. Conversely, advancements in ML such as synthetic data generation, robustness and sensitivity analysis, and out-of-domain testing deserve wider adoption in computational pathology.  In this paper, we propose a set of guidelines for evaluating ML models in pathology which will help researchers create better datasets, metrics, and evaluation setup. We believe this will improve patient outcomes through ML applications in pathology and inform real-world motivated research. 



\subsubsection*{Acknowledgments}
Authors would like to thank Jonathan Rubin (Philips, Amazon) and Maz Abulnaga (MIT) for their inputs.

\bibliography{opinion_paper}

\begin{thebibliography}{48}
\providecommand{\natexlab}[1]{#1}
\providecommand{\url}[1]{\texttt{#1}}
\expandafter\ifx\csname urlstyle\endcsname\relax
  \providecommand{\doi}[1]{doi: #1}\else
  \providecommand{\doi}{doi: \begingroup \urlstyle{rm}\Url}\fi

\bibitem[Aeffner et~al.(2017)Aeffner, Wilson, Martin, Black, Hendriks, Bolon,
  Rudmann, Gianani, Koegler, Krueger, and Young]{10.5858/arpa.2016-0386-RA}
Famke Aeffner, Kristin Wilson, Nathan~T. Martin, Joshua~C. Black, Cris
  L.~Luengo Hendriks, Brad Bolon, Daniel~G. Rudmann, Roberto Gianani, Sally~R.
  Koegler, Joseph Krueger, and G.~Dave Young.
\newblock {The Gold Standard Paradox in Digital Image Analysis: Manual Versus
  Automated Scoring as Ground Truth}.
\newblock \emph{Archives of Pathology \& Laboratory Medicine}, 141\penalty0
  (9):\penalty0 1267--1275, 05 2017.
\newblock ISSN 0003-9985.
\newblock \doi{10.5858/arpa.2016-0386-RA}.
\newblock URL \url{https://doi.org/10.5858/arpa.2016-0386-RA}.

\bibitem[Badgeley et~al.(2019)Badgeley, Zech, Oakden-Rayner, Glicksberg, Liu,
  Gale, McConnell, Percha, Snyder, and Dudley]{badgeley2019deep}
Marcus~A Badgeley, John~R Zech, Luke Oakden-Rayner, Benjamin~S Glicksberg,
  Manway Liu, William Gale, Michael~V McConnell, Bethany Percha, Thomas~M
  Snyder, and Joel~T Dudley.
\newblock Deep learning predicts hip fracture using confounding patient and
  healthcare variables.
\newblock \emph{NPJ digital medicine}, 2\penalty0 (1):\penalty0 1--10, 2019.

\bibitem[Borkowski et~al.(2019)Borkowski, Bui, Thomas, Wilson, DeLand, and
  Mastorides]{borkowski2019lung}
Andrew~A Borkowski, Marilyn~M Bui, L~Brannon Thomas, Catherine~P Wilson,
  Lauren~A DeLand, and Stephen~M Mastorides.
\newblock Lung and colon cancer histopathological image dataset (lc25000).
\newblock \emph{arXiv preprint arXiv:1912.12142}, 2019.

\bibitem[Bulten et~al.(2022)Bulten, Kartasalo, Chen, Str{\"o}m, Pinckaers,
  Nagpal, Cai, Steiner, van Boven, Vink, et~al.]{bulten2022artificial}
Wouter Bulten, Kimmo Kartasalo, Po-Hsuan~Cameron Chen, Peter Str{\"o}m, Hans
  Pinckaers, Kunal Nagpal, Yuannan Cai, David~F Steiner, Hester van Boven,
  Robert Vink, et~al.
\newblock Artificial intelligence for diagnosis and gleason grading of prostate
  cancer: the panda challenge.
\newblock \emph{Nature medicine}, pp.\  1--10, 2022.

\bibitem[Campanella et~al.(2019)Campanella, Hanna, Geneslaw, Miraflor, Werneck
  Krauss~Silva, Busam, Brogi, Reuter, Klimstra, and
  Fuchs]{campanella2019clinical}
Gabriele Campanella, Matthew~G Hanna, Luke Geneslaw, Allen Miraflor, Vitor
  Werneck Krauss~Silva, Klaus~J Busam, Edi Brogi, Victor~E Reuter, David~S
  Klimstra, and Thomas~J Fuchs.
\newblock Clinical-grade computational pathology using weakly supervised deep
  learning on whole slide images.
\newblock \emph{Nature medicine}, 25\penalty0 (8):\penalty0 1301--1309, 2019.

\bibitem[Carrasco-Zevallos et~al.(2021)Carrasco-Zevallos, Taylor-Weiner,
  Pokkalla, Pouryahya, Biddle-Snead, Han, Huss, Juyal, Shanis, Pedawi,
  et~al.]{carrasco2021ai}
Oscar Carrasco-Zevallos, Amaro Taylor-Weiner, Harsha Pokkalla, Maryam
  Pouryahya, Charles Biddle-Snead, Ling Han, Ryan Huss, Dinkar Juyal, Zahil
  Shanis, Aryan Pedawi, et~al.
\newblock Ai-based histologic measurement of nash (aim-nash): A drug
  development tool for assessing clinical trial end points.
\newblock In \emph{JOURNAL OF HEPATOLOGY}, volume~75, pp.\  S254--S254.
  ELSEVIER RADARWEG 29, 1043 NX AMSTERDAM, NETHERLANDS, 2021.

\bibitem[Castro et~al.(2020)Castro, Walker, and Glocker]{castro2020causality}
Daniel~C Castro, Ian Walker, and Ben Glocker.
\newblock Causality matters in medical imaging.
\newblock \emph{Nature Communications}, 11\penalty0 (1):\penalty0 1--10, 2020.

\bibitem[Davison et~al.(2020)Davison, Harrison, Cotter, Alkhouri, Sanyal,
  Edwards, Colca, Iwashita, Koch, and Dittrich]{DAVISON20201322}
Beth~A. Davison, Stephen~A. Harrison, Gad Cotter, Naim Alkhouri, Arun Sanyal,
  Christopher Edwards, Jerry~R. Colca, Julie Iwashita, Gary~G. Koch, and
  Howard~C. Dittrich.
\newblock Suboptimal reliability of liver biopsy evaluation has implications
  for randomized clinical trials.
\newblock \emph{Journal of Hepatology}, 73\penalty0 (6):\penalty0 1322--1332,
  2020.
\newblock ISSN 0168-8278.
\newblock \doi{https://doi.org/10.1016/j.jhep.2020.06.025}.
\newblock URL
  \url{https://www.sciencedirect.com/science/article/pii/S0168827820303998}.

\bibitem[Diao et~al.(2021)Diao, Wang, Chui, Mountain, Gullapally, Srinivasan,
  Mitchell, Glass, Hoffman, Rao, et~al.]{diao2021human}
James~A Diao, Jason~K Wang, Wan~Fung Chui, Victoria Mountain, Sai~Chowdary
  Gullapally, Ramprakash Srinivasan, Richard~N Mitchell, Benjamin Glass, Sara
  Hoffman, Sudha~K Rao, et~al.
\newblock Human-interpretable image features derived from densely mapped cancer
  pathology slides predict diverse molecular phenotypes.
\newblock \emph{Nature communications}, 12\penalty0 (1):\penalty0 1--15, 2021.

\bibitem[Echle et~al.(2021)Echle, Rindtorff, Brinker, Luedde, Pearson, and
  Kather]{echle2021deep}
Amelie Echle, Niklas~Timon Rindtorff, Titus~Josef Brinker, Tom Luedde,
  Alexander~Thomas Pearson, and Jakob~Nikolas Kather.
\newblock Deep learning in cancer pathology: a new generation of clinical
  biomarkers.
\newblock \emph{British journal of cancer}, 124\penalty0 (4):\penalty0
  686--696, 2021.

\bibitem[England \& Cheng(2019)England and Cheng]{england2019artificial}
Joseph~R England and Phillip~M Cheng.
\newblock Artificial intelligence for medical image analysis: a guide for
  authors and reviewers.
\newblock \emph{American journal of roentgenology}, 212\penalty0 (3):\penalty0
  513--519, 2019.

\bibitem[Faryna et~al.(2021)Faryna, van~der Laak, and
  Litjens]{faryna2021tailoring}
Khrystyna Faryna, Jeroen van~der Laak, and Geert Litjens.
\newblock Tailoring automated data augmentation to h\&e-stained histopathology.
\newblock In \emph{Medical Imaging with Deep Learning}, 2021.

\bibitem[FDA(1995)]{fda_robustness}
FDA.
\newblock {Guideline for Industry: Text on Validation of Analytical Procedures
  }.
\newblock \url{https://www.fda.gov/media/71724/download}, 1995.

\bibitem[Food et~al.()Food, Administration, et~al.]{foodguidance}
Food, Drug Administration, et~al.
\newblock Guidance for industry and food and drug administration staff:
  Computer-assisted detection devices applied to radiology images and radiology
  device data-premarket notification [510 (k)] submissions.

\bibitem[Glass et~al.(2021)Glass, Vandenberghe, Chavali, Javed, Rebelatto,
  Sridharan, Elliott, Rao, Montalto, Resnick, et~al.]{glass2021machine}
Benjamin Glass, Michel~Erminio Vandenberghe, Surya~Teja Chavali, Syed~Ashar
  Javed, Marlon Rebelatto, Shamira Sridharan, Hunter Elliott, Sudha Rao,
  Michael Montalto, Murray Resnick, et~al.
\newblock Machine learning models to quantify her2 for real-time tissue image
  analysis in prospective clinical trials., 2021.

\bibitem[Goh et~al.(2017)Goh, Wang, and Wong]{goh2017batch}
Wilson Wen~Bin Goh, Wei Wang, and Limsoon Wong.
\newblock Why batch effects matter in omics data, and how to avoid them.
\newblock \emph{Trends in biotechnology}, 35\penalty0 (6):\penalty0 498--507,
  2017.

\bibitem[Hendrycks \& Dietterich(2019)Hendrycks and
  Dietterich]{hendrycks2019benchmarking}
Dan Hendrycks and Thomas Dietterich.
\newblock Benchmarking neural network robustness to common corruptions and
  perturbations.
\newblock \emph{arXiv preprint arXiv:1903.12261}, 2019.

\bibitem[Hosseini et~al.(2019)Hosseini, Chan, Tse, Tang, Deng, Norouzi,
  Rowsell, Plataniotis, and Damaskinos]{hosseini2019atlas}
Mahdi~S Hosseini, Lyndon Chan, Gabriel Tse, Michael Tang, Jun Deng, Sajad
  Norouzi, Corwyn Rowsell, Konstantinos~N Plataniotis, and Savvas Damaskinos.
\newblock Atlas of digital pathology: A generalized hierarchical histological
  tissue type-annotated database for deep learning.
\newblock In \emph{Proceedings of the IEEE/CVF Conference on Computer Vision
  and Pattern Recognition}, pp.\  11747--11756, 2019.

\bibitem[Howard et~al.(2020)Howard, Dolezal, Kochanny, Schulte, Chen, Heij,
  Huo, Nanda, Olopade, Kather, et~al.]{howard2020_batch}
Frederick~M Howard, James Dolezal, Sara Kochanny, Jefree Schulte, Heather Chen,
  Lara Heij, Dezheng Huo, Rita Nanda, Olufunmilayo~I Olopade, Jakob~N Kather,
  et~al.
\newblock The impact of digital histopathology batch effect on deep learning
  model accuracy and bias.
\newblock \emph{bioRxiv}, 2020.

\bibitem[Jiang et~al.(2020)Jiang, Yang, Wang, Li, and Sun]{jiang2020emerging}
Yahui Jiang, Meng Yang, Shuhao Wang, Xiangchun Li, and Yan Sun.
\newblock Emerging role of deep learning-based artificial intelligence in tumor
  pathology.
\newblock \emph{Cancer communications}, 40\penalty0 (4):\penalty0 154--166,
  2020.

\bibitem[Karimi et~al.(2019)Karimi, Nir, Fazli, Black, Goldenberg, and
  Salcudean]{karimi2019deep}
Davood Karimi, Guy Nir, Ladan Fazli, Peter~C Black, Larry Goldenberg, and
  Septimiu~E Salcudean.
\newblock Deep learning-based gleason grading of prostate cancer from
  histopathology images—role of multiscale decision aggregation and data
  augmentation.
\newblock \emph{IEEE journal of biomedical and health informatics}, 24\penalty0
  (5):\penalty0 1413--1426, 2019.

\bibitem[Kather et~al.(2020)Kather, Heij, Grabsch, Loeffler, Echle, Muti,
  Krause, Niehues, Sommer, Bankhead, et~al.]{kather2020pan}
Jakob~Nikolas Kather, Lara~R Heij, Heike~I Grabsch, Chiara Loeffler, Amelie
  Echle, Hannah~Sophie Muti, Jeremias Krause, Jan~M Niehues, Kai~AJ Sommer,
  Peter Bankhead, et~al.
\newblock Pan-cancer image-based detection of clinically actionable genetic
  alterations.
\newblock \emph{Nature Cancer}, 1\penalty0 (8):\penalty0 789--799, 2020.

\bibitem[Kleiner et~al.(2005)Kleiner, Brunt, Van~Natta, Behling, Contos,
  Cummings, Ferrell, Liu, Torbenson, Unalp-Arida, Yeh, McCullough, Sanyal, and
  Network]{intra-rater-variability}
David~E. Kleiner, Elizabeth~M. Brunt, Mark Van~Natta, Cynthia Behling,
  Melissa~J. Contos, Oscar~W. Cummings, Linda~D. Ferrell, Yao-Chang Liu,
  Michael~S. Torbenson, Aynur Unalp-Arida, Matthew Yeh, Arthur~J. McCullough,
  Arun~J. Sanyal, and Nonalcoholic Steatohepatitis Clinical~Research Network.
\newblock Design and validation of a histological scoring system for
  nonalcoholic fatty liver disease.
\newblock \emph{Hepatology}, 41\penalty0 (6):\penalty0 1313--1321, 2005.
\newblock \doi{https://doi.org/10.1002/hep.20701}.
\newblock URL
  \url{https://aasldpubs.onlinelibrary.wiley.com/doi/abs/10.1002/hep.20701}.

\bibitem[Koh et~al.(2021)Koh, Sagawa, Marklund, Xie, Zhang, Balsubramani, Hu,
  Yasunaga, Phillips, Gao, et~al.]{koh2021wilds}
Pang~Wei Koh, Shiori Sagawa, Henrik Marklund, Sang~Michael Xie, Marvin Zhang,
  Akshay Balsubramani, Weihua Hu, Michihiro Yasunaga, Richard~Lanas Phillips,
  Irena Gao, et~al.
\newblock Wilds: A benchmark of in-the-wild distribution shifts.
\newblock In \emph{International Conference on Machine Learning}, pp.\
  5637--5664. PMLR, 2021.

\bibitem[Komura \& Ishikawa(2018)Komura and Ishikawa]{komura2018machine}
Daisuke Komura and Shumpei Ishikawa.
\newblock Machine learning methods for histopathological image analysis.
\newblock \emph{Computational and structural biotechnology journal},
  16:\penalty0 34--42, 2018.

\bibitem[Koo \& Li(2016)Koo and Li]{koo2016guideline}
Terry~K Koo and Mae~Y Li.
\newblock A guideline of selecting and reporting intraclass correlation
  coefficients for reliability research.
\newblock \emph{Journal of chiropractic medicine}, 15\penalty0 (2):\penalty0
  155--163, 2016.

\bibitem[Larrazabal et~al.(2020)Larrazabal, Nieto, Peterson, Milone, and
  Ferrante]{larrazabal2020gender}
Agostina~J Larrazabal, Nicol{\'a}s Nieto, Victoria Peterson, Diego~H Milone,
  and Enzo Ferrante.
\newblock Gender imbalance in medical imaging datasets produces biased
  classifiers for computer-aided diagnosis.
\newblock \emph{Proceedings of the National Academy of Sciences}, 117\penalty0
  (23):\penalty0 12592--12594, 2020.

\bibitem[Lipsitch et~al.(2010)Lipsitch, Tchetgen, and Cohen]{negativecontrol}
Marc Lipsitch, Eric~Tchetgen Tchetgen, and Ted Cohen.
\newblock Negative controls: a tool for detecting confounding and bias in
  observational studies.
\newblock \emph{Epidemiology (Cambridge, Mass.)}, 21\penalty0 (3):\penalty0
  383, 2010.

\bibitem[Liu et~al.(2017)Liu, Gadepalli, Norouzi, Dahl, Kohlberger, Boyko,
  Venugopalan, Timofeev, Nelson, Corrado, Hipp, Peng, and
  Stumpe]{liu2017detecting}
Yun Liu, Krishna Gadepalli, Mohammad Norouzi, George~E. Dahl, Timo Kohlberger,
  Aleksey Boyko, Subhashini Venugopalan, Aleksei Timofeev, Philip~Q. Nelson,
  Greg~S. Corrado, Jason~D. Hipp, Lily Peng, and Martin~C. Stumpe.
\newblock Detecting cancer metastases on gigapixel pathology images, 2017.

\bibitem[Madabhushi \& Lee(2016)Madabhushi and Lee]{madabhushi2016image}
Anant Madabhushi and George Lee.
\newblock Image analysis and machine learning in digital pathology: Challenges
  and opportunities.
\newblock \emph{Medical image analysis}, 33:\penalty0 170--175, 2016.

\bibitem[Oakden-Rayner et~al.(2020)Oakden-Rayner, Dunnmon, Carneiro, and
  R{\'e}]{oakden2020hidden}
Luke Oakden-Rayner, Jared Dunnmon, Gustavo Carneiro, and Christopher R{\'e}.
\newblock Hidden stratification causes clinically meaningful failures in
  machine learning for medical imaging.
\newblock In \emph{Proceedings of the ACM conference on health, inference, and
  learning}, pp.\  151--159, 2020.

\bibitem[Ozkan et~al.(2016)Ozkan, Eruyar, Cebeci, Memik, Ozcan, and
  Kuskonmaz]{ozkan2016interobserver}
Tayyar~A Ozkan, Ahmet~T Eruyar, Oguz~O Cebeci, Omur Memik, Levent Ozcan, and
  Ibrahim Kuskonmaz.
\newblock Interobserver variability in gleason histological grading of prostate
  cancer.
\newblock \emph{Scandinavian journal of urology}, 50\penalty0 (6):\penalty0
  420--424, 2016.

\bibitem[Park \& Han(2018)Park and Han]{park2018methodologic}
Seong~Ho Park and Kyunghwa Han.
\newblock Methodologic guide for evaluating clinical performance and effect of
  artificial intelligence technology for medical diagnosis and prediction.
\newblock \emph{Radiology}, 286\penalty0 (3):\penalty0 800--809, 2018.

\bibitem[Raciti et~al.(2020)Raciti, Sue, Ceballos, Godrich, Kunz, Kapur,
  Reuter, Grady, Kanan, Klimstra, et~al.]{raciti2020novel}
Patricia Raciti, Jillian Sue, Rodrigo Ceballos, Ran Godrich, Jeremy~D Kunz,
  Supriya Kapur, Victor Reuter, Leo Grady, Christopher Kanan, David~S Klimstra,
  et~al.
\newblock Novel artificial intelligence system increases the detection of
  prostate cancer in whole slide images of core needle biopsies.
\newblock \emph{Modern Pathology}, 33\penalty0 (10):\penalty0 2058--2066, 2020.

\bibitem[Reinke et~al.(2021)Reinke, Eisenmann, Tizabi, Sudre, R{\"a}dsch,
  Antonelli, Arbel, Bakas, Cardoso, Cheplygina, et~al.]{reinke2021common}
Annika Reinke, Matthias Eisenmann, Minu~D Tizabi, Carole~H Sudre, Tim
  R{\"a}dsch, Michela Antonelli, Tal Arbel, Spyridon Bakas, M~Jorge Cardoso,
  Veronika Cheplygina, et~al.
\newblock Common limitations of image processing metrics: A picture story.
\newblock \emph{arXiv preprint arXiv:2104.05642}, 2021.

\bibitem[Schmitt et~al.(2021)Schmitt, Maron, Hekler, Stenzinger, Hauschild,
  Weichenthal, Tiemann, Krahl, Kutzner, Utikal, et~al.]{schmitt2021_batch}
Max Schmitt, Roman~Christoph Maron, Achim Hekler, Albrecht Stenzinger, Axel
  Hauschild, Michael Weichenthal, Markus Tiemann, Dieter Krahl, Heinz Kutzner,
  Jochen~Sven Utikal, et~al.
\newblock Hidden variables in deep learning digital pathology and their
  potential to cause batch effects: prediction model study.
\newblock \emph{Journal of medical Internet research}, 23\penalty0
  (2):\penalty0 e23436, 2021.

\bibitem[Sch{\"o}mig-Markiefka et~al.(2021)Sch{\"o}mig-Markiefka, Pryalukhin,
  Hulla, Bychkov, Fukuoka, Madabhushi, Achter, Nieroda, B{\"u}ttner, Quaas,
  et~al.]{stresstesting}
Birgid Sch{\"o}mig-Markiefka, Alexey Pryalukhin, Wolfgang Hulla, Andrey
  Bychkov, Junya Fukuoka, Anant Madabhushi, Viktor Achter, Lech Nieroda,
  Reinhard B{\"u}ttner, Alexander Quaas, et~al.
\newblock Quality control stress test for deep learning-based diagnostic model
  in digital pathology.
\newblock \emph{Modern Pathology}, 34\penalty0 (12):\penalty0 2098--2108, 2021.

\bibitem[Seyyed-Kalantari et~al.(2020)Seyyed-Kalantari, Liu, McDermott, Chen,
  and Ghassemi]{seyyed2020chexclusion}
Laleh Seyyed-Kalantari, Guanxiong Liu, Matthew McDermott, Irene~Y Chen, and
  Marzyeh Ghassemi.
\newblock Chexclusion: Fairness gaps in deep chest x-ray classifiers.
\newblock In \emph{BIOCOMPUTING 2021: Proceedings of the Pacific Symposium},
  pp.\  232--243. World Scientific, 2020.

\bibitem[Shao et~al.(2021)Shao, Bian, Chen, Wang, Zhang, Ji,
  et~al.]{shao2021transmil}
Zhuchen Shao, Hao Bian, Yang Chen, Yifeng Wang, Jian Zhang, Xiangyang Ji,
  et~al.
\newblock Transmil: Transformer based correlated multiple instance learning for
  whole slide image classification.
\newblock \emph{Advances in Neural Information Processing Systems}, 34, 2021.

\bibitem[Sudharshan et~al.(2019)Sudharshan, Petitjean, Spanhol, Oliveira,
  Heutte, and Honeine]{sudharshan2019multiple}
PJ~Sudharshan, Caroline Petitjean, Fabio Spanhol, Luiz~Eduardo Oliveira,
  Laurent Heutte, and Paul Honeine.
\newblock Multiple instance learning for histopathological breast cancer image
  classification.
\newblock \emph{Expert Systems with Applications}, 117:\penalty0 103--111,
  2019.

\bibitem[Taylor-Weiner et~al.(2021)Taylor-Weiner, Pokkalla, Han, Jia, Huss,
  Chung, Elliott, Glass, Pethia, Carrasco-Zevallos, et~al.]{taylor2021machine}
Amaro Taylor-Weiner, Harsha Pokkalla, Ling Han, Catherine Jia, Ryan Huss,
  Chuhan Chung, Hunter Elliott, Benjamin Glass, Kishalve Pethia, Oscar
  Carrasco-Zevallos, et~al.
\newblock A machine learning approach enables quantitative measurement of liver
  histology and disease monitoring in nash.
\newblock \emph{Hepatology}, 74\penalty0 (1):\penalty0 133--147, 2021.

\bibitem[Tellez et~al.(2019)Tellez, Litjens, B{\'a}ndi, Bulten, Bokhorst,
  Ciompi, and Van Der~Laak]{tellez2019quantifying}
David Tellez, Geert Litjens, P{\'e}ter B{\'a}ndi, Wouter Bulten, John-Melle
  Bokhorst, Francesco Ciompi, and Jeroen Van Der~Laak.
\newblock Quantifying the effects of data augmentation and stain color
  normalization in convolutional neural networks for computational pathology.
\newblock \emph{Medical image analysis}, 58:\penalty0 101544, 2019.

\bibitem[Varoquaux \& Cheplygina(2021)Varoquaux and
  Cheplygina]{varoquaux2021failed}
Ga{\"e}l Varoquaux and Veronika Cheplygina.
\newblock How i failed machine learning in medical imaging--shortcomings and
  recommendations.
\newblock \emph{arXiv preprint arXiv:2103.10292}, 2021.

\bibitem[Veeling et~al.(2018)Veeling, Linmans, Winkens, Cohen, and
  Welling]{veeling2018rotation}
Bastiaan~S Veeling, Jasper Linmans, Jim Winkens, Taco Cohen, and Max Welling.
\newblock Rotation equivariant cnns for digital pathology.
\newblock In \emph{International Conference on Medical image computing and
  computer-assisted intervention}, pp.\  210--218. Springer, 2018.

\bibitem[Waters \& Der(2018)Waters and Der]{waters2018kras}
Andrew~M Waters and Channing~J Der.
\newblock Kras: the critical driver and therapeutic target for pancreatic
  cancer.
\newblock \emph{Cold Spring Harbor perspectives in medicine}, 8\penalty0
  (9):\penalty0 a031435, 2018.

\bibitem[Yagi(2011)]{yagi2011color}
Yukako Yagi.
\newblock Color standardization and optimization in whole slide imaging.
\newblock In \emph{Diagnostic pathology}, volume~6, pp.\  1--12. Springer,
  2011.

\bibitem[Yu \& Eng(2020)Yu and Eng]{yu2020one}
Alice~C Yu and John Eng.
\newblock One algorithm may not fit all: how selection bias affects machine
  learning performance.
\newblock \emph{Radiographics}, 40\penalty0 (7):\penalty0 1932--1937, 2020.

\bibitem[Zech et~al.(2018)Zech, Badgeley, Liu, Costa, Titano, and
  Oermann]{zech2018variable}
John~R Zech, Marcus~A Badgeley, Manway Liu, Anthony~B Costa, Joseph~J Titano,
  and Eric~Karl Oermann.
\newblock Variable generalization performance of a deep learning model to
  detect pneumonia in chest radiographs: a cross-sectional study.
\newblock \emph{PLoS medicine}, 15\penalty0 (11):\penalty0 e1002683, 2018.

\end{thebibliography}
\bibliographystyle{iclr_style}


\end{document}